\documentclass[a4paper,12pt]{article}

\usepackage{graphics,amsmath,amssymb,epsfig,epstopdf}
\usepackage{array}
\usepackage{longtable}
\usepackage{lscape}
\usepackage{pdflscape}
\usepackage{multirow}
\usepackage[utf8]{inputenc}

\begin{document}

\title{
%\begin{flushright}
%{\normalsize IRB-TH-??/08}
%\end{flushright}
%\vspace{2 cm}
\bf Barotropic fluid compatible parametrizations of dark energy}

\author{Dalibor Perkovi\'{c}$^{1,}$\thanks{dalibor.perkovic@zvu.hr}  and Hrvoje \v Stefan\v ci\'c$^{2,}$\thanks{hrvoje.stefancic@unicath.hr}}

%\author{Hrvoje \v Stefan\v ci\'c\thanks{shrvoje@thphys.irb.hr}}

\vspace{3 cm}
\date{
\centering
$^{1}$ University of Applied Health Sciences, Mlinarska street 38, 10000 Zagreb, Croatia \\
\vspace{0.2cm}
$^{2}$ Catholic University of Croatia, Ilica 242, 10000 Zagreb, Croatia }

%\institute{
%  Theoretical Physics Division, Ru\dj er Bo\v{s}kovi\'{c} Institute,
%   P. O. Box 1016, HR-10001 Zagreb, Croatia}

\maketitle

\abstract{Parametrizations of Equation of state parameter as a function of the scale factor or redshift are frequently used in dark energy modeling. The question investigated in this paper is if parametrizations proposed in the literature are compatible with the dark energy being a barotropic fluid. The test of this compatibility is  based on the functional form of the speed of sound squared, which for barotropic fluid dark energy follows directly from the function for the Equation of state parameter. The requirement that the speed of sound squared should be between 0 and speed of light squared provides constraints on model parameters using analytical and numerical methods. It is found that this fundamental requirement eliminates a large number of parametrizations as barotropic fluid dark energy models and puts strong constraints on parameters of other dark energy parametrizations.       }

\vspace{2cm}

\section{Introduction}

\label{intro}

Ever more precise observations of various cosmic phenomena \cite{SNIa1,SNIa2,CMB1,CMB2,BAO,NatureAstronomy} reveal a present state of universe which cannot be understood only in terms of General Relativity and forms of matter known from local physics, such as radiation or baryonic matter. Available observations point to the presently accelerated cosmic expansion, whereas the dynamics at the level of galaxies and clusters of galaxies, among other places, reveals additional gravitational interaction which could be explained by the presence of large quantities of, yet not directly observed, dark matter. The mechanism behind the accelerated cosmic expansion is usually attributed to a cosmic component with the negative pressure, called dark energy (DE). It has been shown that the concept of dark energy can be realized in many different ways such as cosmological constant, dynamical cosmological term, quintessence, phantom energy, k-essence or interacting dark energy \cite{DERev1,DERev2,DERev3,DERev4,DERev5,DERev6}. Numerous alternatives have been proposed to both dark matter and dark energy, frequently as a modification of gravitational interaction at scales from galactic to cosmic \cite{ModGravRev1,ModGravRev2,ModGravRev3}. Yet, even if the effects such as accelerated cosmic expansion or galactic rotation curve dynamics do not originate from cosmic components, concepts of dark matter and dark energy (including their unifications) remain very useful effective concepts. Present observational data reveal a large tension in the value of $H_0$ inferred from low and high redshift measurements assuming the benchmark $\Lambda$CDM model (for a recent review see \cite{Riess}). Some of proposed solutions to this puzzle are nontrivial dynamics of dark energy \cite{NatureAstronomy} and dark matter-dark energy interaction \cite{DiValentino}. 

Presently, various models of dark energy have been proposed that available observational data cannot efficiently discriminate. Physically very distinct DE models can produce very similar global DE evolution and, correspondingly, very similar history of global cosmic expansion. Without a preferred DE model, the fits to observational data have to be performed for a large number of dynamically near-degenerate models. In such a situation a number of researchers have adopted a phenomenological approach of modeling the equation of state (EoS) parameter as a function of the scale factor, $w=w(a)$ (or equivalently of the cosmic redshift, $w=w(z)$) \cite{Efstathiou,Cooray,Chevallier,Linder,Hannestad,Wetterich,Lee,Gong,Nesseris,Ichikawa,Barboza,Liu,Jassal,Ma,Sendra,Feng,Sello,Wei,Zhang,Pantazis,Mamon,Das,Yang}. This approach simplifies the analysis of DE dynamics and allows the analysis of physically interesting $w(a)$ functions. Although such parametrizations may constitute a phenomenological approach of their own to the modeling of dark energy, their main purpose is the simplification of the fits to the observational data. In this way a single simple $w(a)$ parametrization may represent the dynamical behavior of a large number of DE models. Yet, it is important to know to which extent the choice of some parametrization limits its representation by some specific physical model. In particular, it would be interesting to know if some DE models cannot be represented by some $w(a)$ parametrization. In this paper we particularly focus on barotropic fluid models of dark energy and investigate which $w(a)$ parametrizations are compatible with barotropic fluid DE. Here barotropic fluid is understood as a fluid for which the fluid pressure is a function of fluid energy density only. In determining the compatibility we do not employ the comparison of particular parametrizations with the observational data, but rely on fundamental physical constraints on the fluid DE speed of sound.        

If we assume that the dark energy specified by some particular $w(a)$ (or equivalently $w(z)$) parametrization is physically a barotropic fluid, an explicit expression for the barotropic speed of sound squared can be obtained from $w(a)$. Inserting the definition of speed of sound squared for barotropic fluids, $c_s^2 = \frac {d \, p}{d \, \rho}$ and the Equation of State (EoS) parameter $w=\frac{p}{\rho}$ into the continuity equation
\begin{equation}
\label{eq:cont}
d \rho + 3 (\rho+p) \frac{d a}{a} =0 \, ,
\end{equation}
yields the dynamical equation for the EoS parameter
\begin{equation}
\label{eq:wofa}
a \frac{d w}{ d a} = -3 (1+w)(c_s^2-w) \, .
\end{equation}  
This equation can be easily rearranged to obtain the expression for $c_s^2$ in terms of $w$ and $a \frac{d w}{d a}$:
\begin{equation}
\label{eq:cs2}
c_s^2=w - \frac{1}{3(1+w)} a \frac{d w}{d a} \, .
\end{equation}

If $a \frac{d w}{d a}$ can be expressed as a function of $w$, then the speed of sound squared can also be expressed as a function of $w$, i.e. $c_s^2=c_s^2(w)$. This line of modeling has been successfully applied to the description of cosmological constant boundary crossing \cite{mi1} and dark energy-dark matter unification \cite{Caplar,mi2}.

The barotropic fluid speed of sound squared is physically constrained to be nonnegative and not larger than speed of light squared, $c^2$. As we work in system of units where $c=1$, these requirements translate to $0 \le c_s^2 \le 1$. For known $w(a)$ one can obtain $c_s^2(a)$ from (\ref{eq:cs2}) and model parameters for which  $0 \le c_s^2(a) \le 1$ is satisfied for the entire past cosmic expansion, i.e. for the entire $[0,a_0]$ interval. In this way we can select $w(a)$ parametrizations which are suitable for the description of barotropic fluid dark energy as those for which the condition on speed of sound squared is satisfied at least for some model parameters. The allowed region of model parameters is further analyzed if some of its portion corresponds to presently  accelerating component (corresponding to $\rho+3 p <0$). This program, though physically simple, turns out to be quite restrictive for a large number of DE parametrization models.

The paper is organized as follows. The first section brings the introduction and the presentation of the main idea. In the second section we present analytical approach to determination of allowed model parameters and apply it to a one-parameter model and elaborate general methods useful in analytical treatment. In the third section we present numerical approaches to determination of allowed parameter values and apply them to a large number of parametrizations available in the literature. In the following section we discuss the obtained results and finish the paper with conclusions. In the Appendix we bring the analytical solution for the Chevallier-Polarski-Linder (CPL) 
model \cite{Chevallier,Linder}.

\section{Analytical results}

\label{analytical}

The feasibility of constraining the model parameters analytically crucially depends on the form of the $w(a)$ function and constraints can be obtained analytically only in specific cases. Even in cases where the said constraints can be obtained using analytical techniques, the very procedure can be quite involved and the obtained results are not very transparent and informative. Still, analytically tractable cases can be very useful for the verification of more generally applicable numerical approaches and they can provide additional insights that numerical approaches do not provide. As an illustration we describe the analytical procedure for obtaining parameter constraints for a one-parameter model \cite{Gong} and a more general approach suitable for two-parameter models such as the CPL model \cite{Chevallier,Linder}.   

\subsection{$w=w_0 \frac{a}{a_0}$ model}

\label{one_parameter}

Starting from the parametrization \cite{Gong}
\begin{equation}
\label{eq:an_ 1param}
w=w_0 \frac{a}{a_0} \, ,
\end{equation}
from (\ref{eq:cs2}) one obtains
\begin{equation}
\label{eq:cs2_1param}
c_s^2=w_0 \frac{a}{a_0} \frac{2 + 3 w_0 \frac{a}{a_0} }{3(1 + w_0 \frac{a}{a_0})} = w \frac{2+3w}{3(1+w)} = w \left(1-\frac{1}{3(1+w)} \right) \, .
\end{equation}
The first and second derivative of $c_s^2$ with respect to $w$ are:
\begin{equation}
\label{eq:dcdw_1param}
\frac{d c_s^2}{d w} = 1 - \frac{1}{3(1+w)^2} \, ,
\end{equation}
\begin{equation}
\label{eq:d2cdw2_1param}
\frac{d^2 c_s^2}{d w^2} = \frac{2}{3} \frac{1}{(1+w)^3} \, .
\end{equation}
Stationary points of $c_s^2$ are at $w_1=-1-\frac{1}{\sqrt{3}}$ and  $w_2=-1+\frac{1}{\sqrt{3}}$ where at $w_1$ the $c_s^2$ has a maximum and at $w_2$ it has a minimum. Both of these stationary points correspond to negative values of $w$. At $w=-1$ there is a singularity in $c_s^2$ corresponding to the crossing of the cosmological constant boundary.

During the cosmic expansion the EoS parameter (\ref{eq:an_ 1param}) does not change its sign, i.e. as $a$ increases from 0 to $a_0$, $w$ changes from 0 to $w_0$ (increases for positive $w_0$ and decreases for negative $w_0$). 

For negative $w_0$ (negative vaules of $w$), expression (\ref{eq:cs2_1param}) reveals that $c_s^2$ is negative for $-\frac{2}{3}<w<0$. Therefore, if $w_0 > -\frac{2}{3}$, $c_s^2$ is negative during the entire cosmic past, whereas if $w_0 < -\frac{2}{3}$, $c_s^2$ is negative for $a$ from 0 to some  finite $a_* < a_0$. In both cases the condition $c_s^2 \ge 0$ is violated in the cosmic past and $w_0 < 0$ does not correspond to viable fluid model of dark energy.

For positive $w_0$, there are no stationary points in the interval of $w$ between 0 and $w_0$. The expression (\ref{eq:dcdw_1param}) reveals that $\frac{d c_s^2}{d w}$ is positive in the interval $(0,w_0)$ and $c_s^2$ is a growing function of $w$. The expression (\ref{eq:cs2_1param}) shows that $c_s^2$ is positive in the entire considered interval. Therefore, to fulfill the requirement $ 0 \le  c_s^2 \le 1$ at the entire interval, it suffices to require that $c_s^2(w_0) \le 1$. Straightforward calculation shows that this is satisfied for
\begin{equation}
\label{eq:w0_1param}
0 \le w_0 \le \frac{1+\sqrt{37}}{6} \, .
\end{equation}  

As this allowed paramter range corresponds to $w \ge 0$, the parametrization from \cite{Gong} is clearly unsuitable as a model of barotropic fluid dark energy.

\subsection{General analytical approach}

\label{general_analytical}

From the condition $0 \le c_s^2 \le 1$ it is possible to obtain general analytical constraints between $a \frac{d w}{d a}$ and $w$ \cite{LinderScherrer}. In particular, for $1+w >0$ the condition translates to
\begin{equation}
\label{eq:anwgt-1}
-(1-w)(1+w) \le \frac{a}{3} \frac{d w }{d a} \le w(1+w) \, ,
\end{equation}
whereas for $1+w < 0$, the condition results in
\begin{equation}
\label{eq:anwlt-1}
w(1+w) \le \frac{a}{3} \frac{d w }{d a} \le -(1-w)(1+w) \, .
\end{equation}

Furthermore, it is straightforward to show that the condition $0 \le c_s^2 \le 1$ is equivalent to
\begin{equation}
\label{eq:cs2_onecondition}
c_s^2 (c_s^2-1) \le 0 \, ,
\end{equation}
which can also be presented in the form
\begin{equation}
\label{eq:fg}
\frac{f(a) g(a)}{9 (1+w)^2} \le 0 \, .
\end{equation}
Here, bearing in mind that $w=w(a)$,
\begin{equation}
\label{eq:f}
f(a)= 3 w (1+w) - a \frac{d w}{d a}\, 
\end{equation}
and
\begin{equation}
\label{eq:g}
g(a) = 3 (w-1) (1+w) - a \frac{d w}{d a}\, .
\end{equation}
From (\ref{eq:fg}) one can determine the regions of allowed model parameters as those for which $f(a) g(a) \le 0$ for all $ a \in [0, a_0]$. In the  parametrizations for which $ a \frac{d w}{d a}$ can be expressed as a function of $w$, this condition then reads as $f(w) g(w) \le 0$. For a majority of parametrizations the determination of the regions of allowed model parameters cannot be pursued analitically. In the Appendix we systematically apply this approach for the CPL model $w=w_0+w_1(1-\frac{a}{a_0})$, introduced in \cite{Chevallier,Linder}. Although in this case both $f$ and $g$ are (only) quadratic functions of $w$, the analytical calculations require examination of a number of various cases.

\section{Numerical results}

\label{numerical}

\begin{figure}[!t]
\centering
\includegraphics[scale=0.40]{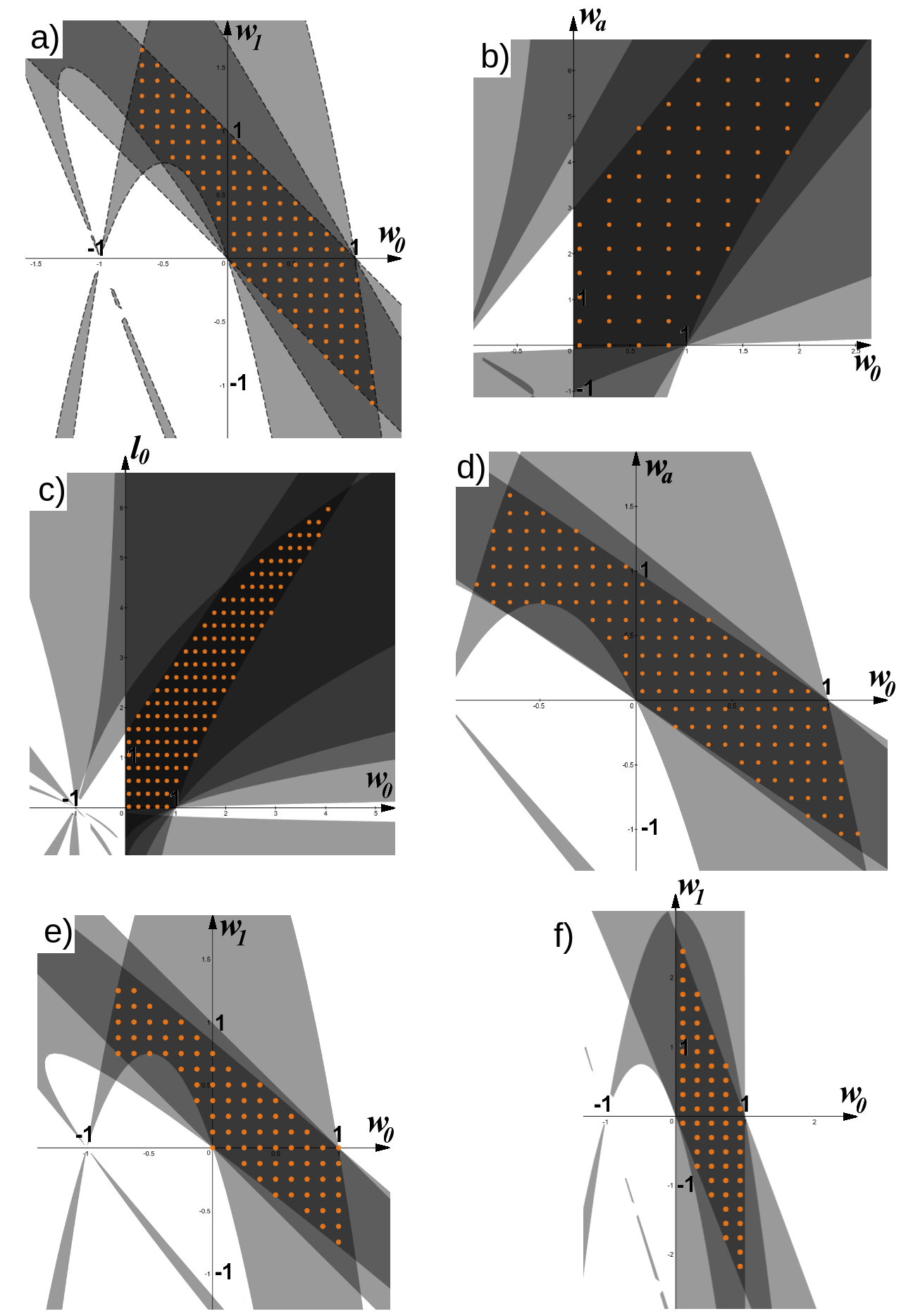}
\caption{The allowed parameter regions for two-parameter models \cite{Chevallier,Linder} in plot a), \cite{Wetterich} (model 1) in plot b), \cite{Wetterich} (model 2) in plot c),  \cite{Pantazis} (model 1)  in plot d), \cite{Barboza} in plot e) and  \cite{Yang} in plot f). In all plots the symbol $w_0$ on the axis denotes the present value of the $w(a)$ function, whereas the other symbols refer to parameters in the corresponding $w(a)$ parametrizations.
\label{fig1}}
\end{figure}

\begin{figure}[!t]
\centering
\includegraphics[scale=0.35]{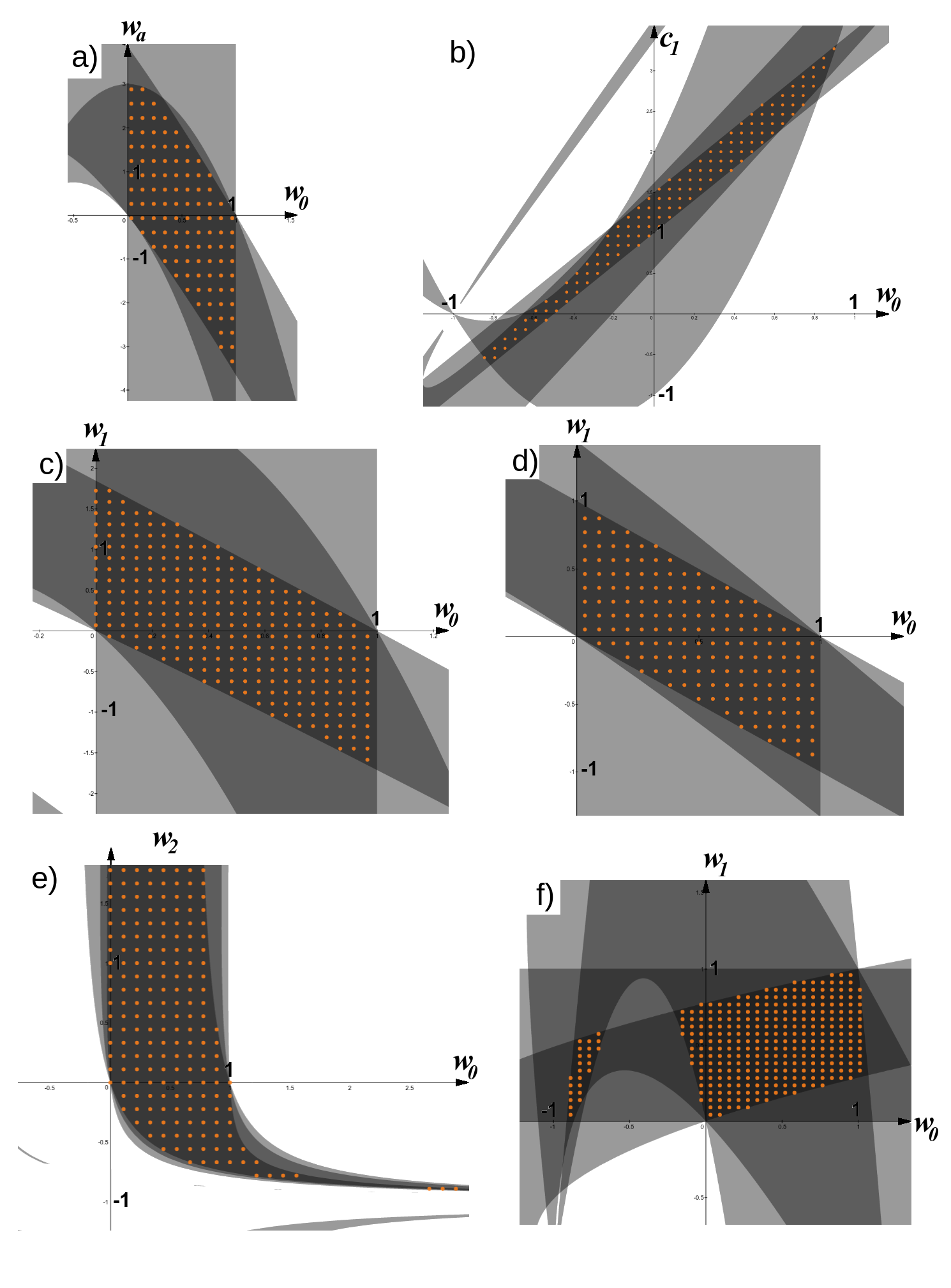}
\caption{The allowed parameter regions for two-parameter models  \cite{Jassal} in plot a), \cite{Sendra} in plot b), \cite{Feng} (model 1) in plot c), \cite{Feng} (model 2) in plot d), \cite{Zhang} in plot e) and three-parameter model \cite{Ichikawa} for $a_*=a_0/3$ in plot f).  In all plots the symbol $w_0$ on the axis denotes the present value of the $w(a)$ function, whereas the other symbols refer to parameters in the corresponding $w(a)$ parametrizations.
\label{fig2}}
\end{figure}

\begin{figure}[!t]
\centering
\includegraphics[scale=0.35]{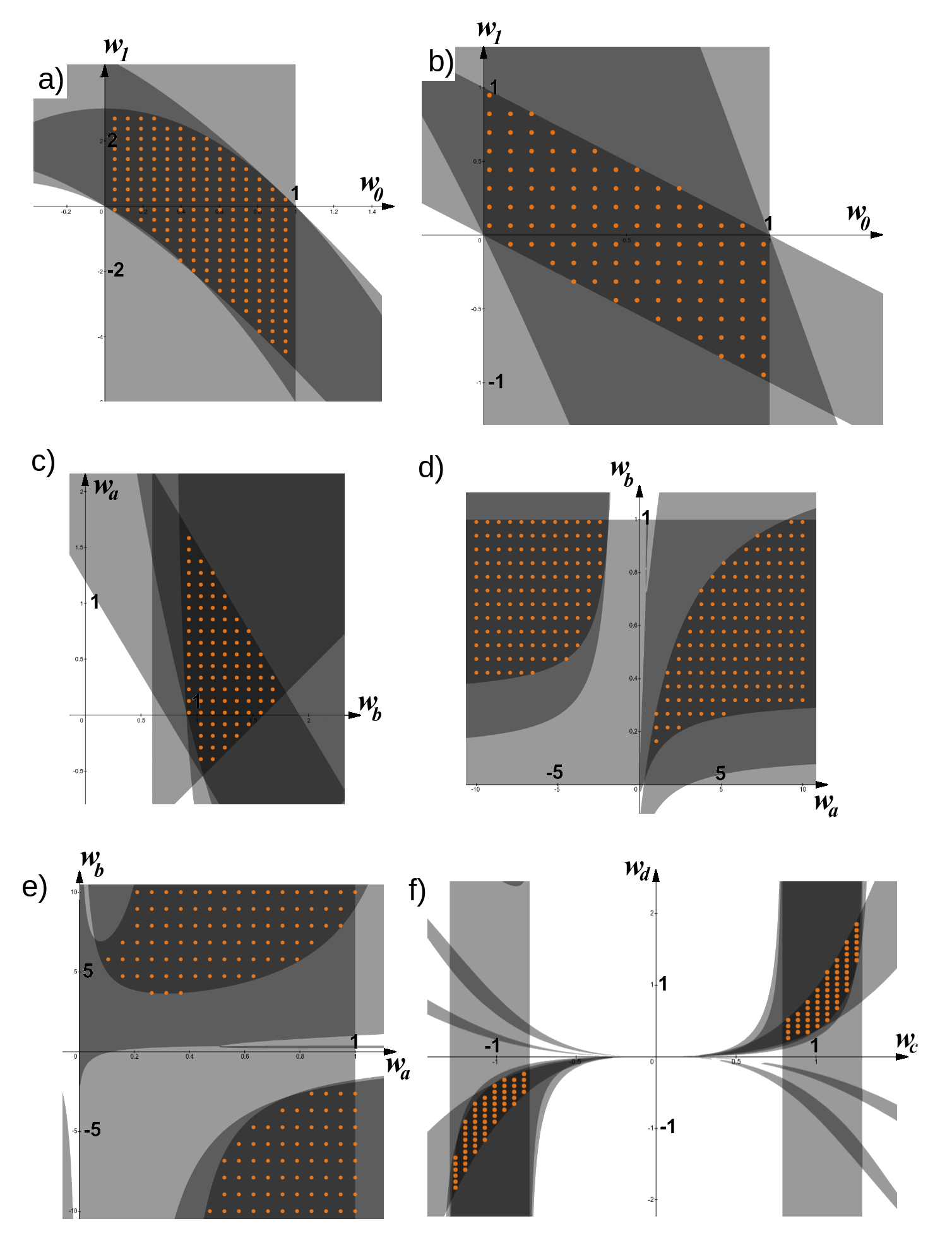}
\caption{The allowed parameter regions for three-parameter models \cite{Liu} (model 1) for $n=3$ in plot a), \cite{Pantazis} (model 2) and \cite{Liu} for $n=3$ in plot b) and 
\cite{Wei} for $w_0=-0.6$ in plot c) and four-parameter models \cite{Hannestad} for $p=1$, $w_0 = -0.6$ and 
$(\frac{a_s}{a_0})^p = \frac{w_b}{w_a} \frac{w_a - w_0}{w_0 - w_b}$ in plot d),  \cite{Lee} for $p=1$, $w_0 = -0.6$ and 
$(\frac{a_c}{a_0})^p = \frac{w_a w_b - w_0}{w_0 - w_a}$ in plot e) and \cite{Das} $w_0=-0.6$ and $w_b=-1$ in plot f).  In all plots the symbol $w_0$ on the axis denotes the present value of the $w(a)$ function, whereas the other symbols refer to parameters in the corresponding $w(a)$ parametrizations.
\label{fig3}}
\end{figure}

The compatibility of a particular parametrization with the barotropic fluid DE can in general be established only numerically. For the studied parametrizations we determine the allowed region of the model parametric space using two approaches and present the results in Table \ref{tab:1}. For models which have a nonvanishing parameter region corresponding to $0 \le c_s^2 \le 1$ for the entire cosmic past these regions are depicted in Figures \ref{fig1} to \ref{fig3}. 

The graphs in Figures \ref{fig1} to \ref{fig3} were made combining two methods: 
\begin{enumerate}
\item Shaded areas: An analytical solution was rearranged to put one parameter on each axis ($x$ and $y$) and solution space for discrete values of $a$ was graphed (0, $a_0$ and one or two points in the middle; usually, but not always, $a/a_0=0.5$, depending which value created a better illustration of the effect). The intersection of these three (or four) areas approximates a solution for the whole range of $a \in [0, a_0 ]$\footnote{We acknowledge the use of Desmos graphics tool https://www.desmos.com/}. 
\item Dots: A numerical solution was computed with two parameters laid on $x$ and $y$ axes. The condition $0 \le c_s^2 \le 1$ was tested at a set of scale factor $a$ values, where the values in the set were chosen for each model to cover the entire cosmic past, but also to produce the best coverage of the allowed parameter region. In most cases, one hundred values of $a$ were calculated using $a = e^{i/10} a_0$ with $i$ taking integer values from -100 to 0. This produced an array of values for $a$ more dense near $a = 0$ and more spread out near $a = a_0$. A point was placed on the graph for all values of the two parameters where $c_s^2 (a) \in [0, 1] $ was true for {\it all} values of $a \in [0, a_0]$.
\end{enumerate}

Following the initial test for $c_s^2 (a) \in [0, 1] $, this condition was loosened to only $c_s^2 (a) \ge 0$ without the $c_s^2 (a) \le 1$ condition. As expected, the $(w_0, w_1)$ space expanded with the following restrictions:

\begin{enumerate}
\item If the favourable area under the condition $c_s^2 (a) \in [0, 1] $ existed only for values $w_0 > 0$, the expansion happened solely in the direction $w_0 > 0$.
\item If the favourable area existed for values $w_0 < 0 $, but did not reach $w_0 = -1$, it expanded towards $w_0 = -1$, but did not cross to $w_0 < -1$.

\end{enumerate} 

\begin{figure}[!t]
\centering
\includegraphics[scale=0.35]{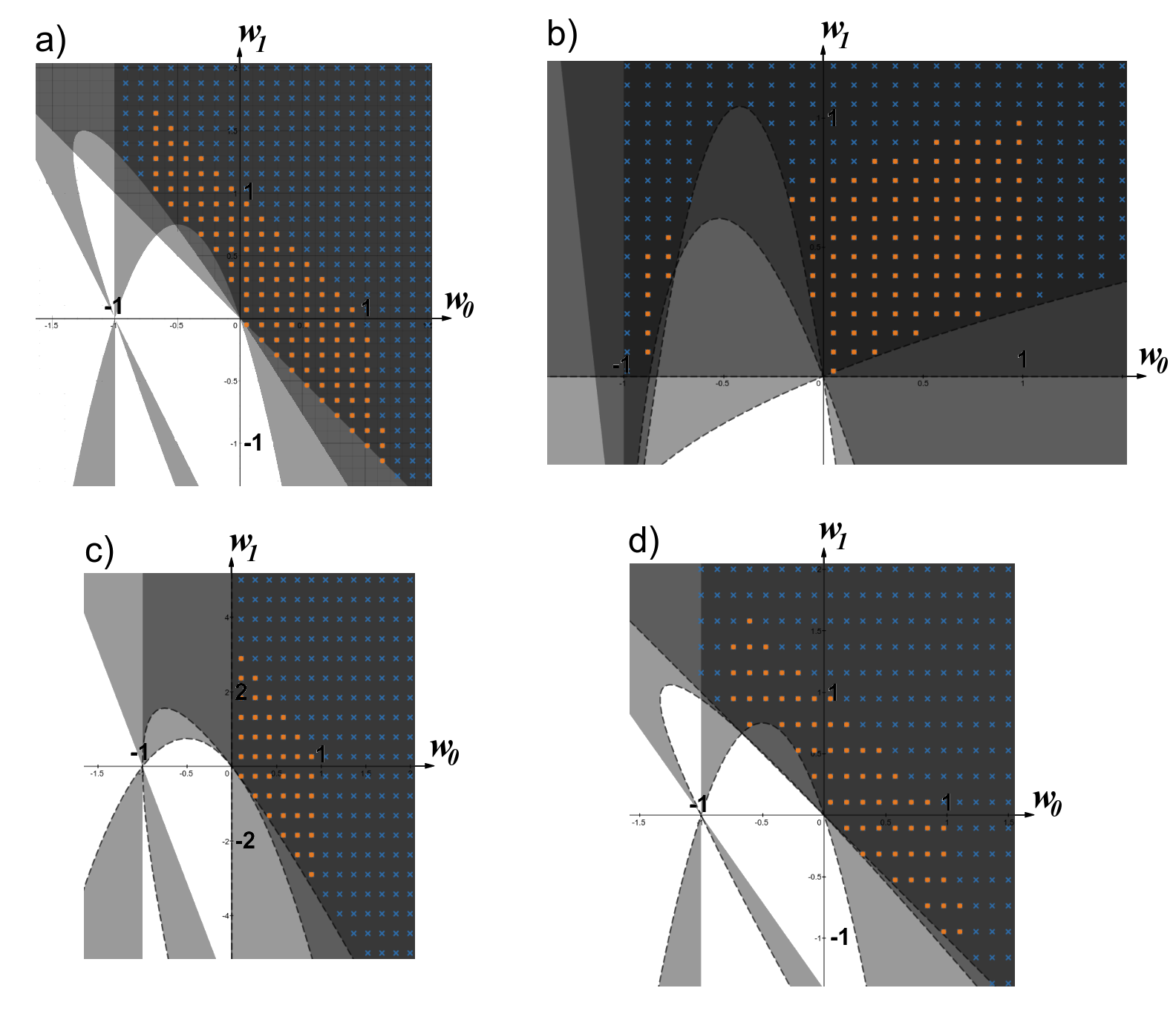}
\caption{The allowed paremeter regions for conditions $0 \le c_s^2 \le 1$ (dots) and $0 \le c_s^2$ (dots + crosses) for models: \cite{Chevallier,Linder} in plot a), \cite{Ichikawa} for $a_* = a_0 / 3$ in plot b), \cite{Jassal} in plot c) and \cite{Pantazis} (model 1) for $n = 3$ in plot d).  In all plots the symbol $w_0$ on the axis denotes the present value of the $w(a)$ function, whereas the other symbols refer to parameters in the corresponding $w(a)$ parametrizations.
\label{fig4}}
\end{figure}

For a number of parametrizations the allowed parameter regions for the condition $c_s^2 \ge 0$  are presented in Figure \ref{fig4}.

\begin{longtable}{ |m{7cm}|m{1.5cm}|m{2cm}|m{3.5cm}| }

 \hline
 
 Formula for $w(a)$ & Ref. & $0 \le c_s^2 \le 1$ 
 \newline 
 for 
 \newline $0 \le a \le a_0$ & Allowed
\newline parameter 
 \newline region \\
 \hline 
 \hline

 \multicolumn{4}{|c|}{One-parameter models} \\
 \hline
 \hline

 $w(a) = \frac{w_0}{a_0} a  $  &  \cite{Gong}  & yes & $w_0 \in [0, \frac{1}{6}+\frac{\sqrt{37}}{6}] $ \\ 
 \hline

  $ w(a) = \frac{w_0}{a_0}  a e^{1 - \frac{a}{a_0}}$ & \cite{Gong}  & yes & $w_0 \in [0, 1] $  \\ 
\hline

  $ w(a) = -1 + \frac{\frac{2\alpha}{3}\frac{a}{a_0}(1-\frac{a}{a_0})}{1 + \alpha (1 - \frac{a}{a_0})^2}$ & \cite{Mamon} & no  & \\
\hline
\hline

 \multicolumn{4}{|c|}{Two-parameter models} \\
 \hline
 \hline

$w(a) = w_0 + w_a ln\frac{a}{a_0} $ & \cite{Efstathiou} & no & \\ 
\hline

 $ w(a) = w_0 + w_a (\frac{a_0}{a} - 1)$  & \cite{Cooray}  & no  & \\ 
 \hline
 
  $w(a)= w_0 + w_1 (1- \frac{a}{a_0})$
 & \cite{Chevallier} 
\cite{Linder}  & yes & Fig 1a\\ 
\hline

 $ w(a) = \frac{w_0}{1-w_a ln \frac{a}{a_0}}$ & \cite{Wetterich}   & yes & Fig 1b\\ 
\hline

 $ w(a) = \frac{w_0}{\big( 1 - l_0 ln\frac{a}{a_0})^2} $
  & \cite{Wetterich}  & yes & Fig 1c \\ 
\hline

"Sqrt model", 
$w(a) = w_0 + w_a \frac{\frac{a_0}{a} - 1}{\sqrt{1 + (\frac{a_0}{a} - 1)^2}} $ & \cite{Pantazis}    & yes & Fig 1d \\ \hline

$w(a) = w_0 + w_1 \frac{ \frac{a_0}{a} \big(\frac{a_0}{a} - 1\big) }{1 +  \big(\frac{a_0}{a} - 1\big)^2}$ &  \cite{Barboza}   & yes & Fig 1e \\ 
 \hline

$ w(a) = w_0 - w_1 \frac{a}{a_0} ln \frac{a}{a_0} $ &  \cite{Yang}  & yes & Fig 1f \\ 

\hline

 $w(a) = w_0 + w_a \frac{a}{a_0} (1-\frac{a}{a_0})$ & \cite{Jassal}   & yes & Fig 2a \\ 
 \hline

 $ w(z) = w_0 + w_a \big(\frac{ln(2+z)}{1+z} - ln2 \big) $ & \cite{Ma}  & no & \\ 
 \hline

 $ w(a) = w_0 + w_1 \Big( \frac{a}{a_0} \sin \big( \frac{a_0}{a} \big) - \sin1 \Big) $  & \cite{Ma} & no & \\ 
 \hline

$w(a) = -1 + c_1 \ \left(2-\frac{a}{a_0}\right)+c_2\ \left(2-\frac{a}{a_0}\right)^2$ &
  \cite{Sendra} & yes & Fig 2b for $c_2=w_0-c_1+1$ \\ 
\hline

 $w(a)=w_0+w_1 \frac{\frac{a_0}{a} - 1}{1+(\frac{a_0}{a} - 1)^2}$ &  \cite{Feng}  & yes & Fig 2c\\ \hline
 
 $w(a)=w_0+w_1 \frac{(\frac{a_0}{a} - 1)^2}{1+(\frac{a_0}{a} - 1)^2}$ &  \cite{Feng} & yes & Fig 2d \\ 
 \hline

 $ w(z) = w_0 + w_a \big(\frac{ln \sqrt{1 + z^2} - ln\sqrt{z}}{1+z} + ln \sqrt{2} \big) $ & \cite{Sello}  & no \footnote{$c_s^2 < 0$ for small values of $a$ ($\frac{a}{a_0}< 10^{-8}$ or less, depending on parameters)} & \\
 \hline

$w(a) = w_1 + \frac{1}{3} \frac{\frac{a_0}{a}}{w_2 + \frac{a_0}{a}}$
 &   \cite{Zhang} & yes & Fig 2e for $ w_1 = w_0 - \frac{1}{3} \frac{1}{w_2 + 1}$\\  
\hline

 \hline \hline

 \multicolumn{4}{|c|}{Three-parameter models} \\
 \hline
  \hline

 $ w(a) = w_0 + w_a \big(\frac{ln(\xi+\frac{a_0}{a})}{\xi+\frac{a_0}{a}-1} - \frac{ln(\xi+1)}{\xi} \big) $ & \cite{Sello} & no \footnote{$c_s^2 < 0$ for small values of $a$ ($\frac{a}{a_0}< 10^{-8}$ or less, depending on parameters)} & \\ 
 \hline

$w(a) = w_0 + \frac{w_1 - w_0}{z_*} (\frac{a_0}{a}-1), a>a_*$ \newline
$w(a)=w_1, a<a_*$  &   \cite{Ichikawa}  & yes & Fig 2f for $a_* = a_0/3$\\ 
\hline

$w(a)=w_0+w_1 \left(1-\frac{a}{a_0}\right) \Big(\frac{a}{a_0}\Big)^{n-1} $ &   \cite{Liu}  & yes & Fig 3a for $n=3$\\

\hline

 "Generalised CPL" \newline
$w(a)= w_0 + w_1 (1- \frac{a}{a_0})^n$ & 
\cite{Liu}\cite{Pantazis}  & yes & Fig 3b for $n=3$ \\ \hline

$ w(z) = \frac {a_1 + 3(\Omega_{m0} - 1) - 2a_1 z - a_2 (z^2 + 2z -2)} {3 (1 - \Omega_{m0} + a_1 z + 2a_2 z + a_2 z^2)} $ &  \cite{Nesseris} & no & \\ 
\hline

$ w = \frac{w_0 + (w_0 + w_a)(\frac{a_0}{a}-1)}{1 + (1 + w_b)(\frac{a_0}{a}-1)}$ & \cite{Wei}  & yes & Fig 3c for $w_0 = -0.6$ \\ 
\hline

$w(a) = \frac{w_0 + w_1 \ln \frac{a}{a_0}}{1 + w_2 \ln \frac{a}{a_0}} $ & \cite{Wei}  & no & \\ 
\hline

 \hline

 \multicolumn{4}{|c|}{Four-parameter models} \\
 \hline
 \hline

$ w(a) = w_a w_b \frac{(\frac{a}{a_0})^p + (\frac{a_s}{a_0})^p}{w_b(\frac{a}{a_0})^p + w_a(\frac{a_s}{a_0})^p}$ &  \cite{Hannestad}  & yes & Fig 3d for $p = 1$, $w_0 = -0.6$
$(\frac{a_s}{a_0})^p = \frac{w_b}{w_a} \frac{w_a - w_0}{w_0 - w_b}$\\ 
\hline

$w(a) = w_a \frac{w_b (\frac{a}{a_0})^p + (\frac{a_c}{a_0})^p}{(\frac{a}{a_0})^p + (\frac{a_c}{a_0})^p}$ & \cite{Lee}  & yes & Fig 3e for $p = 1$, $w_0 = -0.6$
$(\frac{a_c}{a_0})^p = \frac{w_a w_b - w_0}{w_0 - w_a}$\\ 
\hline

%$w(a) = \frac{w_0 + w_1 \ln \frac{a}{a_0}}{1 + w_2 \ln \frac{a}{a_0}} $ & \cite{Wei}  & no & \\ 
%\hline

$ w(a) = w_a + \frac{w_b}{\Big(w_c + w_d(\frac{a_0}{a} - 1)\Big)^2}$ & \cite{Das}  & yes & Fig 3f  for $w_0 = -0.6$, $w_a = w_0 - \frac{w_b}{w_c^2}$ and $w_b = -1$\\ 

 \hline
 
  \multicolumn{4}{c}{} \\

 \caption{Overview of studied models, their $w(a)$ (or $w(z)$) parametrizations and allowed regions of parameters.}
 \label{tab:1}

\end{longtable}

Plots in Figures \ref{fig1} and \ref{fig2}  present the two-parameter models with nonvanishing allowed regions of parameters. In all these plots the parameter at the $x$ axis is $w_0$, corresponding to the present value of the EoS parameter $w(a_0)$ (or $w(z=0)$). It is interesting to observe which models allow $w_0 < -1/3$ values (which can in principle serve as accelerating components). In Figure \ref{fig1} this condition is satisfied for models in plots a) \cite{Chevallier,Linder}, d) \cite{Pantazis} and e) \cite{Barboza}. In Figure \ref{fig2} presently accelerating component is possible for the model \cite{Sendra} in plot b) and the model \cite{Ichikawa} in plot f).

For three-parameter and four-parameter models, presented in Figure \ref{fig3}, all models presented may describe barotropic fluid dark energy. The plots demonstrate that for the selected values of $w_0=w(a_0)$ there are nonvanishing allowed regions of other model parameters. This fact shows that the capability of these models to represent baryonic DE model is not the result of some contrived combinations of model parameters, but a generic feature of these models.

\section{Discussion and conclusions}

In total, of the three one-parameter models studied in this paper, two have a nonvanishing parameter space consistent with the requirement $0 \le c_s^2 \le 1$, but none of the allowed parameters corresponds to the presently accelerating component. For the 16 two-parameter models we analyzed, 11 of them have a nonvanishing parameter space consistent with   $0 \le c_s^2 \le 1$, but only four of these can describe a presently accelerating component. For ten of the three-parameter and four-parameter models, seven of them satisfy the requirement  $0 \le c_s^2 \le 1$ and are compatible with the barotropic fluid dark energy.

One can observe that virtually none of  studied models allows values of $w_0$ very close to $-1$. This fact may be attributed to the term $\sim \frac{1}{1+w}$ in the expression (\ref{eq:cs2}) for $c_s^2$ which diverges when $w \rightarrow -1$, which correspondingly blows up the value of $c_s^2$. A natural question arising is which of two requirements ($c_s^2 \ge 0$ or $c_s^2 \le 1$) is responsible for such a behavior. Indeed, if the $c_s \le 1$ requirement is relaxed, the allowed values of $w_0$ are much closer to $-1$, as is evident from Figure \ref{fig4}. 

As one could expect, the larger the number of model parameters, it is easier to find their combination for which the model may be represented as a barotropic fluid DE in the entire cosmic past. For the studied single parameter models  it is found that none of them can satisfy the requirement on $c_s^2$ and presently have a sufficiently negative $w$. For two-parameter models, only four out of 16 models are capable of representing the barotropic fluid dark energy. For three-parameter and four-parameter models seven of the ten studied models fit the requirement of baryonic fluid dark energy.

It is important to notice (possibly even somewhat surprising) that these strong restrictions have been obtained on purely theoretical, but fundamental  grounds. The parametrizations $w(a)$ that are found to be able to represent a barotropic fluid DE by the studied requirements on $c_s^2$ still need to be compared against the available observational data which will  further constrain the parametric space obtained in this paper. 

The results of this paper indicate that the suitability of a phenomenological parametrization $w(a)$ to describe some physically motivated dark energy model need not come automatically. Internal theoretical features of a chosen physical DE models may constrain, or even eliminate some phenomenological parametrizations. Possible situations in which this kind of argumentation might be applicable, apart from the barotropic fluid DE models, comprise k-essence DE models or effective DE description of modified gravity theories.

\section*{Appendix}

In this Appendix we elaborate the analytical determination of the allowed region of model parameters for the CPL model \cite{Chevallier,Linder}, starting from the formalism developed in section \ref{general_analytical}. For the CPL model the functions $f(w)$ and $g(w)$ are second order polynomials in $w$ in particular 
$$c_s^2 \ge 0 : f(w)=3w^2 + 2w + w_0 + w_1 $$
and
$$c_s^2 \le 1 : g(w)=3w^2 - w + w_0 + w_1 - 3 \, .$$
Their zeros are
$$f(w)=0 \; : w_\pm^0 = -\frac{1}{3} \pm \frac{\sqrt{1-3(w_0 + w_1)}}{3} \, ,$$
for $1-3(w_0 + w_1) \ge 0$  and 
$$g(w)=0 \; : w_\pm^1 = \frac{1}{6} \pm \frac{\sqrt{37-12(w_0 + w_1)}}{6} \, .$$
for $37-12(w_0 + w_1) \ge 0$.

The next step in the procedure is determination of conditions for both $f(w)$ and $g(w)$ being $\ge 0$ or $\le 0$. We denote respective conditions by the letter $S$ and corresponding indices as they will translate into regions of the $w_0 - w_1$ plane.

The condition
$$ \boxed{f(w) \ge 0}$$
can be realized in two subcases. If the function $f(w)$ has no zeros the condition is 
$$S_{01}: 1-3(w_0 + w_1) < 0 \Rightarrow w \in \langle -\infty , +\infty \rangle \, , $$
whereas if $f(w)$ has zeros the condition reads
$$S_{02}: 1-3(w_0 + w_1) \ge 0 \Rightarrow w \in \langle -\infty , w_-^0 ] \cup [ w_+^0 , +\infty \rangle \, .$$
The requirement 
$$ \boxed{f(w) \le 0}$$
translates to
$$S_{03}: 1-3(w_0 + w_1) \ge 0 \Rightarrow w \in [ w_-^0 , w_+^0] \, .$$

The condition
$$ \boxed{g(w) \ge 0}$$
has two subcases. If the function $g(w)$ has no zeros the condition leads to
$$S_{11}: 37-12(w_0 + w_1) < 0  \Rightarrow w \in \langle -\infty , +\infty \rangle \, ,$$
while if $g(w)$ has zeros the condition translates to
$$S_{12}: 37-12(w_0 + w_1) \ge 0 \Rightarrow w \in \langle -\infty , w_-^1 ] \cup [ w_+^1 , +\infty \rangle \, .$$
On the other hand, the condition
$$ \boxed{g(w) \le 0}$$
leads to
$$S_{13}: 37-12(w_0 + w_1) \ge 0 \Rightarrow w \in [ w_-^1 , w_+^1] \, .$$

The additional requirement is that all values that $w(a)$ acquires in the interval $[0, a_0]$ (interval between $w(0)=w_0+w_1$ and $w(a_0)=w_0$) have to be contained in the allowed intervals of $w$ obtained in the consideration of $f(w)$ and $g(w)$ functions. Further elaboration depends on the sign of $w_1$. We assume $w_1 \neq 0$, since $w_1=0$ leads to a trivial case $w=w_0$.
$$\boxed{\boxed{S^+: w_1>0}}$$
The interval of variation of $w$ is
$$ \boxed{w \in [ w_0, w_0 + w_1 ]}$$
Individual conditions for nonnegative $f(w)$ then read
$$S_{01}^+ : 1-3(w_0 + w_1) < 0$$
and
$$S_{02}^+ : [w_0, w_0 + w_1] \subseteq \langle -\infty , w_-^0] \cup [w_+^0 , +\infty \rangle $$
which leads to
$$ 1-3(w_0 + w_1) \ge 0 \cap  ( (w_0 + w_1 \le w_-^0) \cup (w_0 \ge w_+^0) ) \, . $$
The condition for nonpositive $f(w)$ reads
$$S_{03}^+ : [w_0, w_0 + w_1] \subseteq [ w_-^0 , w_+^0] $$
resulting in 
$$ 1-3(w_0 + w_1) \ge 0 \cap (w_0 \ge w_-^0) \cap (w_0 + w_1 \le w_+^0) \, .$$
The conditions for nonnegative $g(w)$ are
$$S_{11}^+ : 37-12(w_0 + w_1) < 0 $$
and
$$S_{12}^+ : [w_0, w_0 + w_1] \subseteq \langle -\infty , w_-^1] \cup [w_+^1 , +\infty \rangle $$
which leads to 
$$37-12(w_0 + w_1) \ge 0 \cap ( (w_0 + w_1 \le w_-^1) \cup (w_0 \ge w_+^1) ) \, .$$
The condition for nonpositive $g(w)$ gives
$$S_{13}^+ : [w_0, w_0 + w_1] \subseteq  [ w_-^1 , w_+^1] $$
resulting in the condition
$$ 37-12(w_0 + w_1) \ge 0 \cap (w_0 \ge w_-^1) \cap (w_0 + w_1 \le w_+^1) \, .$$

$$\boxed{\boxed{S^-: w_1<0}}$$
The interval of variation of $w$ is
$$ \boxed{w \in [ w_0 + w_1, w_0 ]} \, .$$
The conditions for nonnegative $f(w)$ are
$$S_{01}^-  : 1-3(w_0 + w_1) < 0$$
and
$$S_{02}^- : [w_0 + w_1, w_0] \subseteq \langle -\infty , w_-^0] \cup [w_+^0 , +\infty \rangle $$
resulting in
$$ 1-3(w_0 + w_1) \ge 0 \cap( (w_0  \le w_-^0) \cup (w_0 + w_1 \ge w_+^0) )\, . $$
The condition for nonpositive $f(w)$ leads to
$$S_{03}^- : [w_0 + w_1, w_0] \subseteq [w_-^0 , w_+^0] $$
which  reads
$$1-3(w_0 + w_1) \ge 0 \cap (w_0 + w_1 \ge w_-^0) \cap (w_0 \le w_+^0) \, .$$
The conditions for nonnegative $g(w)$ are
$$S_{11}^- : 37-12(w_0 + w_1) < 0 $$
and
$$S_{12}^- : [w_0 + w_1, w_0] \subseteq \langle -\infty , w_-^1] \cup [w_+^1 , +\infty \rangle$$
resulting in
$$ 37-12(w_0 + w_1) \ge 0 \cap ( (w_0 \le w_-^1) \cup (w_0 + w_1 \ge  w_+^1)) \, .$$
The condition for nonpositive $g(w)$ is
$$S_{13}^- : [w_0 + w_1, w_0] \subseteq  [w_-^1 , w_+^1]$$
leading to 
$$ 37-12(w_0 + w_1) \ge 0 \cap (w_0 + w_1 \ge w_-^1) \cap (w_0 \le w_+^1) \, .$$
 
 \newpage
 
The overall condition determining the allowed region of model parameters can be expressed as 

$$ S^+ \cap  \Big( \big[ ( S_{01}^+ \cup S_{02}^+ ) \cap S_{13}^+\big] \cup \big[   ( S_{11}^+  \cup S_{12}^+ ) \cap S_{03}^+  \big] \Big) $$
$$\cup$$
$$ S^-  \cap \Big( \big[ ( S_{01}^- \cup S_{02}^- ) \cap S_{13}^-\big] \cup \big[   ( S_{11}^-  \cup S_{12}^- )  \cap S_{03}^-  \big] \Big) \, .$$

However, this can be simplified. Let us investigate some parts of this solution. Firstly, let us consider
$$ \tilde{S}^+ = S^+ \cap \big[     ( S_{11}^+  \cup S_{12}^+ ) \cap S_{03}^+  \big],  $$
which can be written as 
$$(S^+ \cap S_{11}^+ \cap S_{03}^+) \cup  (S^+ \cap S_{12a}^+ \cap S_{03}^+) \cup (S^+ \cap S_{12b}^+ \cap S_{03}^+) \, ,$$
where 
$$S_{12a}^+ = 37-12(w_0+w_1) \ge 0 \cap w_0 + w_1 \le \frac{1-\sqrt{37-12(w_0+w_1)}}{6} \, ,$$
$$S_{12b}^+ = 37-12(w_0+w_1) \ge 0 \cap w_0 \ge \frac{1+\sqrt{37-12(w_0+w_1)}}{6}\, .$$

We immediately see that $w_0+w_1>\frac{37}{12}$ from $S_{11}^+$ is incompatible with $w_0 + w_1 \le \frac{1}{3}$ from $S_{03}^+$. This leads to 
$$S^+ \cap S_{11}^+ \cap S_{03}^+ = \emptyset \, .$$

Next,  it can be shown that $S_{12}^+a$ is equvalent to
$$ w_0+w_1 \le -1 \, .$$
Now, $S^+_{12a}$ means $w_1 > 0$ which also leads to $w_0 + w_1 > w_0$. Combining it with the second part of $S_{03}^+$ yields
$$3(w_0 + w_1) + 1 > - \sqrt{1-3(w_0 + w_1)}$$
which can be shown to be equivalent to
$$w_0 + w_1 > -1.$$
This is an obvious contradiction, so
$$S^+ \cap S_{12a}^+ \cap S_{03}^+ = \emptyset \, .$$

Finally,  it can easily be shown that the third part of $S_{03}^+$ is equivalent to
$$ w_0 + w_1 \le 0 \, , $$
meaning:
$$w_0 < 0.$$
However, the second part of $S_{12}^b$ says that
$$ w_0 \ge \frac{1}{6} + \frac{\sqrt{37-12(w_0 + w_1)}}{6} \, .$$
so this also ends in contradiction, meaning that
$$S^+ \cap S_{12b}^+ \cap S_{03}^+ = \emptyset \, .$$
These results together lead to  $\tilde{S}^+ = \emptyset$.

Next, we use the equivalent procedure to examine
$$ \tilde{S}^- =  S^- \cap \big[   ( S_{11}^-  \cup S_{12}^- )  \cap S_{03}^-  \big],  $$
which can be written as 
$$(S^- \cap S_{11}^- \cap S_{03}^-) \cup  (S^- \cap S_{12a}^- \cap S_{03}^-) \cup (S^- \cap S_{12b}^- \cap S_{03}^-) \, ,$$
where 
$$S_{12a}^- = 37-12(w_0+w_1) \ge 0 \cap w_0  \le \frac{1-\sqrt{37-12(w_0+w_1)}}{6} \, ,$$
$$S_{12b}^- = 37-12(w_0+w_1) \ge 0 \cap w_0 + w_1 \ge \frac{1+\sqrt{37-12(w_0+w_1)}}{6}\, .$$

 As in the previous section, $S_{11}^-$ leads to $w_0 + w_1 > \frac{37}{12}$ which is in contradiction with $w_0 + w_1 \le \frac{1}{3}$ from $S_{03}^-$ so that

 $$S^- \cap S_{11}^- \cap S_{03}^- = \emptyset \, .$$

Regarding the second term,  the second part of $S_{03}^-$ leads to $w_0+w_1 \ge -1$, meaning also $w_0 > -1$. However, this leads to contradiction with the second part of $S_{12a}^-$:
$$w_0  \le \frac{1-\sqrt{37-12(w_0+w_1)}}{6}$$
resulting in 
$$S^- \cap S_{12a}^- \cap S_{03}^- = \emptyset \, .$$
Finally, the second part of $S_{12b}^-$ leads to $w_0+w_1 \ge 1$ which is in contradiction with $w_0+w_1 \le 1/3$, resulting in
 $$S^- \cap S_{12b}^- \cap S_{03}^- = \emptyset \, .$$

These results together lead to  $\tilde{S}^- = \emptyset$.

Therefore,  the overall allowed region of model parameters is

$$\Big( S^+ \cap \big[ ( S_{01}^+ \cup S_{02}^+ ) \cap S_{13}^+\big] \Big) 
\cup
\Big( S^- \cap \big[ ( S_{01}^- \cup S_{02}^- ) \cap S_{13}^-\big] \Big)  .$$

This set of conditions fully corresponds with the numerically obtained allowed parameter region presented in plot a) of Figure \ref{fig1}.

\end{document}